\documentclass[conference]{IEEEtran}

\usepackage{cite}
\usepackage{amsmath,amssymb,amsfonts}
\usepackage{algorithmic}
\usepackage{graphicx}
\usepackage{textcomp}
\usepackage{xcolor}
\usepackage{braket}
\usepackage{booktabs}
\usepackage{multirow}
\usepackage{url}
\usepackage[letterpaper, left=0.625in, right=0.625in, top=0.75in, bottom=1.00in]{geometry}
\def\BibTeX{{\rm B\kern-.05em{\sc i\kern-.025em b}\kern-.08em
    T\kern-.1667em\lower.7ex\hbox{E}\kern-.125emX}}
\begin{document}

\title{VQC--ZTI: Variational Quantum Control for Zero Trust Protection of the Tactile Internet}

\author{\IEEEauthorblockN{1\textsuperscript{st} Mubassir Serneabat Sudipto}
\IEEEauthorblockA{\textit{Electrical and Computer Engineering} \\
\textit{Iowa State University}\\
Ames, Iowa, USA \\
msudipto@iastate.edu}
\and
\IEEEauthorblockN{2\textsuperscript{nd} Shakil Ahmed}
\IEEEauthorblockA{\textit{Computer Science, College of Computing} \\
\textit{Grand Valley State University}\\
Allendale, Michigan, USA \\
ahmeshak@gvsu.edu}
\and
\IEEEauthorblockN{3\textsuperscript{rd} Ashfaq Khokhar}
\IEEEauthorblockA{\textit{Carl R. Ice College of Engineering} \\
\textit{Kansas State University}\\
Manhattan, Kansas, USA \\
akhokhar@ksu.edu}
}

\maketitle

\begin{abstract}
Tactile Internet services couple cyber events directly to physical actuation, so security decisions must improve risk discrimination without perturbing the control path. This paper presents VQC--ZTI, a split-plane Variational Quantum Classifier framework for zero-trust protection of Tactile Internet services, in which an off-path VQC analyzes encrypted-flow telemetry while an on-path policy engine applies cached deterministic grant, restrict, step-up, and deny actions. By decoupling anomaly scoring from enforcement, VQC--ZTI preserves predictable control behavior and allows detector sensitivity and policy aggressiveness to be tuned independently. We evaluate the framework on CESNET-derived aggregated traffic using random, entity-group, and temporal holdouts with a hybrid PyTorch--PennyLane implementation. The full-hybrid Quantum Neural Network achieves mean areas under the receiver operating characteristic curve of 0.9981, 0.9974, and 0.9941 and reduces the false-positive rate relative to ExtraTrees by 44.6\%, 49.6\%, and 67.9\%, respectively. A representative component-timing decomposition further illustrates that batched VQC scoring remains in the asynchronous evidence path rather than the immediate enforcement path.
\end{abstract}


\section{Introduction}\label{sec:introduction}

The Tactile Internet (TI) extends communication to real-time sensing, actuation, teleoperation, and haptic interaction \cite{fettweis2014tactile,maier2016tactile,promwongsa2021tactile}. By coupling network decisions with physical behavior, TI services impose stringent latency, reliability, and control-stability requirements \cite{durisi2016shortpackets,popovski2018urllc}. Encrypted traffic, changing endpoints, multi-slice operation, and resource-constrained gateways also shift security monitoring toward payload-independent flow aggregates and adaptive behavioral baselines \cite{sommer2010outside,ring2019datasets}. The challenge is therefore to use uncertain security evidence without disrupting latency-critical control.
Zero Trust (ZT) addresses this challenge through continuous verification, least privilege, and micro-segmentation \cite{rose2020zta,chandramouli2023ztaModel}. However, directly translating every detector output into an access-control action may cause unnecessary authentication, restrictions, denials, and policy churn. TI-oriented ZT must therefore support informative security analysis while preserving lightweight, deterministic enforcement.
Variational Quantum Circuits (VQCs) and Quantum Neural Networks (QNNs) provide compact differentiable models for structured classical data \cite{biamonte2017qml,preskill2018nisq,mitarai2018qcl,schuld2020circuit}. 

In VQC--ZTI, classical encrypted-flow features are encoded through parameterized rotations, while entangling gates model correlations in the quantum representation. This does not imply that the traffic contains inherent quantum correlations or that the model achieves quantum advantage. The implementation instead uses noiseless analytic simulation as a hybrid quantum--classical evidence model. This paper presents VQC--ZTI, a split-plane architecture that performs hybrid VQC anomaly scoring on mirrored telemetry outside the latency-critical path. In parallel, Policy Enforcement Points (PEPs) apply cached deterministic actions on-path. A Policy Decision Point (PDP) combines anomaly evidence with contextual risk to select grant, restrict, step-up, or deny actions. 

Separate detector and policy thresholds allow anomaly sensitivity and enforcement aggressiveness to be calibrated independently. Unlike earlier QNN-enhanced ZT work for broader next-generation networks \cite{ahmed2025qnnztf}, VQC--ZTI focuses on TI operation and evaluates its evidence model under random, entity-group, and temporal holdouts.
The main contributions are:
\begin{itemize}
    \item \textit{Split-plane VQC--ZTI architecture:} An off-path hybrid VQC performs anomaly scoring, while PEPs provide deterministic cached enforcement on the latency-critical path.
    \item \textit{Decoupled detection and policy control:} Separate detector and policy thresholds enable independent calibration of anomaly sensitivity and enforcement aggressiveness without placing probabilistic inference in the actuation loop.
    \item \textit{TI-oriented empirical evaluation:} A hybrid PyTorch--PennyLane model is evaluated on 4,875 CESNET-derived encrypted-flow records under random, entity-group, and temporal holdouts and compared with classical baselines and quantum-model ablations. Quantile-derived labels are treated as statistical anomalies rather than verified intrusions.
\end{itemize}
Section~\ref{sec:system_model} presents the system model, architecture, and risk logic; Section~\ref{sec:qnn_solution} describes the QNN solution; Section~\ref{sec:perform_eval} reports the evaluation; and Section~\ref{sec:conclusions} concludes the paper.

\section{Proposed System Framework}\label{sec:system_model}

VQC--ZTI is a split-plane framework for zero-trust (ZT) protection of Tactile Internet (TI) services. It improves anomaly-evidence generation without placing probabilistic quantum inference in the latency-critical control path. The evidence plane asynchronously processes mirrored encrypted-flow telemetry through feature preparation, quantum encoding, VQC inference, classical post-processing, risk fusion, and policy computation. The enforcement plane contains Policy Enforcement Points (PEPs) that apply locally cached deterministic policies. Thus, delayed scoring may postpone a future policy update but does not delay the current control or actuation decision.

\subsection{ZT--TI System Model}\label{sec:zt-ti_system_model}

We model the ZT-enabled TI environment as a directed graph
\begin{equation}
G=(V,E),
\label{eq:network_graph}
\end{equation}
where $V$ represents users, devices, gateways, services, controllers, and enforcement entities, and $E$ represents their communication and control links. The security-enabled subgraph is
\begin{equation}
G^{Q}=(V^{Q},E^{Q}),
\qquad
V^{Q}\subseteq V,\quad E^{Q}\subseteq E,
\label{eq:security_graph}
\end{equation}
where $V^{Q}$ contains entities supporting telemetry collection, off-path scoring, risk computation, policy distribution, or enforcement, and $E^{Q}$ contains the corresponding signaling and mirrored-telemetry links.
The framework operates through two logically separated planes. In the evidence plane, encrypted-flow statistics are mirrored without payload inspection and processed by the hybrid VQC to produce anomaly scores. The Policy Decision Point (PDP) combines these scores with identity, device, and segment context and distributes the resulting policies to the appropriate PEPs. In the enforcement plane, each PEP performs a deterministic lookup against its cached policy and immediately applies the selected action.
\begin{figure}[ht]
    \centering
    \includegraphics[width=0.96\linewidth]
    {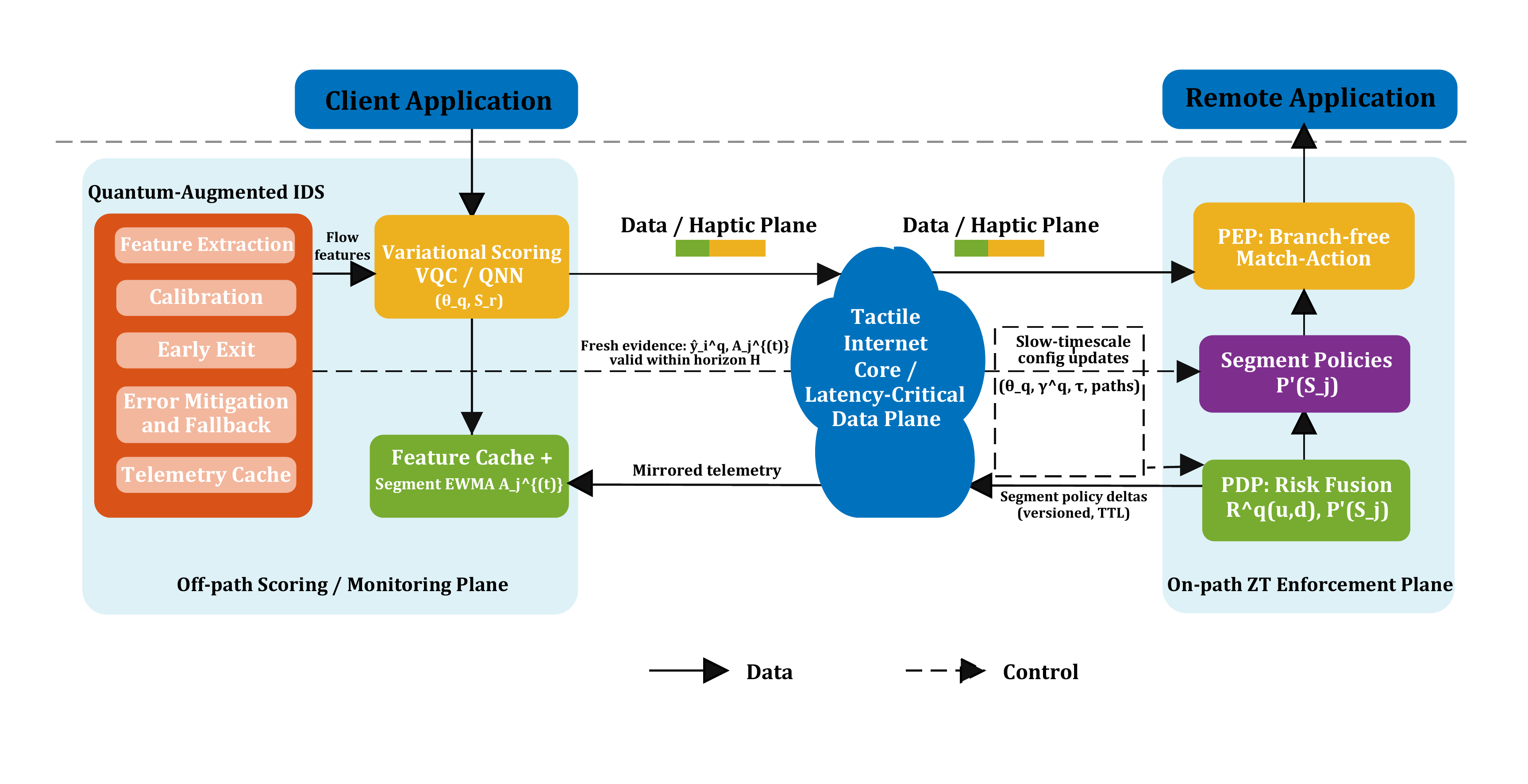}
    \caption{Split-plane VQC--ZTI architecture with off-path evidence generation and on-path deterministic enforcement.}
    \label{fig:vqc_zti_sys_frame}
\end{figure}
Fig.~\ref{fig:vqc_zti_sys_frame} illustrates the separation between the evidence and enforcement planes. In the off-path evidence plane, the VQC mirrors and processes encrypted-flow telemetry to generate anomaly scores. These scores are combined with identity, device, and micro-segment context through risk fusion, after which the PDP computes and distributes updated policies to the relevant PEPs. In the on-path enforcement plane, each PEP performs a lightweight lookup of its locally cached policy and immediately applies the corresponding action grant, restrict, step-up, or deny before the request reaches the protected TI service. Thus, mirrored telemetry influences subsequent policy updates, while VQC inference remains outside the current request path and does not add quantum-processing delay to the immediate enforcement decision.

\subsection{Split-Plane Latency Model}
\label{sec:split_plane_latency}

Let $B_{\mathrm{TI}}$ denote the maximum latency budget for a critical TI transaction. Because VQC scoring and policy computation occur asynchronously, the transaction latency is
\begin{equation}
T_{\mathrm{crit}}
=
T_{\mathrm{comm}}
+
T_{\mathrm{PEP}}
+
T_{\mathrm{service}},
\label{eq:critical_latency}
\end{equation}
where the terms denote communication, cached-policy lookup and enforcement, and protected-service processing delays, respectively. The critical path must satisfy
\begin{equation}
T_{\mathrm{crit}}\leq B_{\mathrm{TI}}.
\label{eq:budget}
\end{equation}
The asynchronous evidence latency is modeled separately as
\begin{equation}
\begin{split}
T_{\mathrm{evid}}
={}&
T_{\mathrm{feat}}
+
T_{\mathrm{enc}}
+
T_{\mathrm{VQC}}
+
T_{\mathrm{head}}\\
&+
T_{\mathrm{PDP}}
+
T_{\mathrm{dist}},
\end{split}
\label{eq:evidence_latency}
\end{equation}
where the terms represent feature preparation, quantum encoding, VQC execution, classical-head inference, PDP computation, and policy distribution. Because this processing is off-path, $T_{\mathrm{evid}}$ is not added to $T_{\mathrm{crit}}$; instead, it determines how quickly telemetry can update a cached policy. For a maximum policy-update interval $T_{\mathrm{fresh}}$,
\begin{equation}
T_{\mathrm{evid}}\leq T_{\mathrm{fresh}}.
\label{eq:policy_freshness}
\end{equation}
This freshness condition differs from the per-transaction requirement in \eqref{eq:budget}. Its value depends on the monitoring interval, application requirements, and deployment policy.

\subsection{Quantum Score Generation and Risk Fusion}
\label{sec:qscore_generation_and_risk_fusion}

For each flow $i$, the evidence plane derives a payload-independent feature vector 
\begin{equation} 
\mathbf{x}_i\in\mathbb{R}^{F},
\label{eq:feature_vector}
\end{equation}
where $F$ denotes the number of input features; in the evaluated
implementation, $F=12$. The hybrid model produces
\begin{equation}
\hat{y}_i^{q}
=
\mathcal{M}_{q}
\left(
\mathbf{x}_i;
\boldsymbol{\theta}_{q}
\right),
\label{eq:qscore2}
\end{equation}
where $\mathcal{M}_{q}$ is the hybrid quantum--classical model, $\boldsymbol{\theta}_{q}$ contains its trainable parameters, $\hat{y}_i^{q}\in[0,1]$ is the anomaly score. This score indicates evidence of a statistical anomaly under the evaluated labeling procedure, not a calibrated probability of malicious activity.
A detector threshold $\gamma^{q}$ produces a binary indication:
\begin{equation}
\begin{aligned}
\hat{y}_i^{q}>\gamma^{q}
&\Rightarrow f_i\ \text{is flagged as anomalous},\\
\hat{y}_i^{q}\leq\gamma^{q}
&\Rightarrow f_i\ \text{is treated as normal},
\end{aligned}
\label{eq:gamma}
\end{equation}
where $f_i$ denotes the observed flow and $\gamma^{q}\in[0,1]$ is the detector threshold. The threshold controls detector sensitivity but does not directly determine enforcement.
At the micro-segment level, VQC--ZTI maintains an Exponentially Weighted Moving Average (EWMA) posture score:
\begin{equation}
\begin{split}
A_j^{(t)}
={}&
(1-\alpha)A_j^{(t-1)}\\
&+
\alpha
\frac{1}{|\mathcal{F}_j^{(t)}|}
\sum_{i\in\mathcal{F}_j^{(t)}}
\mathbf{1}
\left[
\hat{y}_i^{q}>\gamma^{q}
\right],
\end{split}
\label{eq:seg-ewma}
\end{equation}
where $\alpha\in(0,1]$ is the smoothing factor, $\mathcal{F}_j^{(t)}$ is the set of flows observed in segment $j$ during interval $t$, and $\mathbf{1}[\cdot]$ is the indicator function. This aggregation limits the influence of isolated variations, while allowing persistent anomalies to affect the segment's posture. The posture score satisfies $A_j^{(t)}\in[0,1]$ and is initialized as $A_j^{(0)}=0$. If $\mathcal{F}_j^{(t)}=\varnothing$, the previous score is retained, i.e., $A_j^{(t)}=A_j^{(t-1)}$, avoiding division by zero.
The PDP combines the model output with identity, device, and segment context:
\begin{equation}
R^{q}(u,d)
=
\mathcal{F}_{q}
\left(
c_u,
c_d,
\mathbf{x}_i,
\hat{y}_i^{q},
A_j^{(t)}
\right),
\label{eq:risk2}
\end{equation}
where $R^{q}(u,d)\in[0,1]$ is the contextual risk score for user $u$ and device $d$, $c_u$ and $c_d$ represent their context, and $\mathcal{F}_{q}$ is the risk-fusion function. Here, flow $i$ is associated with user $u$, device $d$, and micro-segment $j$. This policy-oriented score is not interpreted as a calibrated probability of malicious activity.

\subsection{Risk-Adaptive Policy Enforcement}
\label{sec:risk_adaptive_enforcement}

The detector threshold $\gamma^{q}$ determines whether evidence is reported as anomalous, whereas the ordered policy thresholds
\begin{equation}
\boldsymbol{\tau}
=
(\tau_1,\tau_2,\tau_3),
\qquad
0\leq\tau_1<\tau_2<\tau_3\leq1,
\label{eq:policy_thresholds}
\end{equation}
map fused risk to an enforcement action:
\begin{equation}
\pi\!\left(R^{q}\right)
=
\begin{cases}
\mathrm{grant},
& R^{q}<\tau_1,\\
\mathrm{restrict},
& \tau_1\leq R^{q}<\tau_2,\\
\mathrm{step\mbox{-}up},
& \tau_2\leq R^{q}<\tau_3,\\
\mathrm{deny},
& R^{q}\geq\tau_3.
\end{cases}
\label{eq:policy_mapping}
\end{equation}
where $\pi(\cdot)$ denotes the PDP policy-mapping function.
Grant permits the requested operation; restrict reduces privileges or tightens micro-segmentation; step-up requires additional verification; and deny blocks the request. Separating $\gamma^{q}$ from $\boldsymbol{\tau}$ allows detector sensitivity and enforcement aggressiveness to be tuned independently, preventing every indication from triggering a disruptive response.
The PDP distributes each updated policy to the relevant PEP, which applies the cached action to subsequent requests. If the evidence plane becomes temporarily unavailable, enforcement continues using the most recent valid policy according to configured expiration and fail-safe rules. The quantum backend is therefore not a hard dependency of immediate TI transactions.
This work focuses on hybrid anomaly-evidence generation, contextual risk fusion, and ZT enforcement. Quantum-network routing, entanglement provisioning, secure-capacity optimization, and reinforcement-learning-based resource allocation are outside the scope of the reported evaluation.

\section{Proposed QNN-Based Solution Approach}
\label{sec:qnn_solution}

The VQC--ZTI evidence plane uses a hybrid quantum--classical model to convert payload-independent encrypted-flow statistics into continuous anomaly scores. Its pipeline comprises feature preprocessing, classical embedding, angle encoding, a shallow variational quantum circuit (VQC), and a classical classification head trained end-to-end using PyTorch and PennyLane. The resulting score feeds the risk-fusion and policy mechanisms in Section~\ref{sec:system_model}; VQC inference remains outside the immediate enforcement path. Hereafter, QNN denotes the complete hybrid architecture, whereas the VQC denotes its parameterized quantum layer.

\subsection{Feature Preparation and Classical Embedding}
\label{sec:feature_preparation}

Let the processed dataset be
\begin{equation}
\mathcal{D}
=
\left\{
\left(
\mathbf{x}_i,y_i
\right)
\right\}_{i=1}^{N},
\qquad
\mathbf{x}_i\in\mathbb{R}^{12},
\quad
y_i\in\{0,1\},
\label{eq:dataset}
\end{equation}
where $N=4{,}875$, $\mathbf{x}_i$ contains 12 payload-independent features derived from CESNET encrypted-traffic aggregates, and $y_i$ is a quantile-derived statistical-anomaly label rather than a verified attack annotation. Robust feature scaling and quantile-derived statistical labeling are performed during dataset preprocessing before construction of the evaluation holdouts.
The resulting processed features are subsequently partitioned according to the random, entity-group, and temporal evaluation protocols. The processed vector is then mapped by
\begin{equation}
\mathbf{z}_i
=
f_{\mathrm{emb}}
\left(
\mathbf{x}_i;
\boldsymbol{\phi}
\right),
\qquad
\mathbf{z}_i\in\mathbb{R}^{12},
\label{eq:classical_embedding}
\end{equation}
where $f_{\mathrm{emb}}(\cdot)$ is the trainable classical embedder with parameters $\boldsymbol{\phi}$. It produces one input per qubit; variants without the embedder encode the processed features directly.

\subsection{Quantum Feature Encoding}
\label{sec:quantum_feature_encoding}

The VQC uses $n_q=12$ qubits initialized in the all-zero state:
\begin{equation}
\ket{\psi_0}
=
\ket{0}^{\otimes n_q},
\label{eq:initial_quantum_state}
\end{equation}
where $\otimes$ denotes the tensor product and $n_q$ is the number of qubits.
Each embedded feature is then encoded as a single-qubit rotation:
\begin{equation}
U_{\mathrm{enc}}
\left(
\mathbf{z}_i
\right)
=
\bigotimes_{k=1}^{n_q}
R_y
\left(
z_{i,k}
\right),
\label{eq:angle_encoding}
\end{equation}
where $z_{i,k}$ denotes the $k$th component of the embedded feature vector $\mathbf{z}_i$, and $R_y(\cdot)$ denotes a rotation about the Pauli-$Y$ axis.
The rotation is defined as
\begin{equation}
R_y(\vartheta)
=
\exp
\left(
-\mathrm{i}\frac{\vartheta}{2}Y
\right),
\label{eq:ry_rotation}
\end{equation}
where $\mathrm{i}=\sqrt{-1}$, $\vartheta$ is the rotation angle, and $Y$ is the Pauli-$Y$ operator.
Applying the encoding operation to the initial state gives
\begin{equation}
\ket{\psi_{\mathrm{enc}}(\mathbf{z}_i)}
=
U_{\mathrm{enc}}
\left(
\mathbf{z}_i
\right)
\ket{\psi_0},
\label{eq:encoded_state}
\end{equation}
where $\ket{\psi_{\mathrm{enc}}(\mathbf{z}_i)}$ denotes the quantum state encoding record $i$. This mapping represents classical features through quantum-state rotations; it does not imply intrinsic quantum properties in the traffic.

\subsection{Variational Quantum Circuit}
\label{sec:variational_quantum_circuit}

The circuit applies a shallow ansatz with $L=2$ variational layers, each containing trainable single-qubit rotations and nearest-neighbor entangling operations:
\begin{equation}
U_{\mathrm{VQC}}
\left(
\boldsymbol{\theta}
\right)
=
\prod_{\ell=1}^{L}
U_{\mathrm{ent}}^{(\ell)}
U_{\mathrm{rot}}^{(\ell)}
\left(
\boldsymbol{\theta}^{(\ell)}
\right),
\qquad
L=2,
\label{eq:vqc_unitary}
\end{equation}
where $\boldsymbol{\theta}$ contains the trainable circuit parameters.
In layer $\ell$, $U_{\mathrm{rot}}^{(\ell)}$ applies a three-parameter single-qubit rotation to each qubit, while $U_{\mathrm{ent}}^{(\ell)}$ applies CNOT gates between consecutive qubits, i.e., $1\!\rightarrow\!2,2\!\rightarrow\!3,\ldots, (n_q-1)\!\rightarrow\!n_q$. The resulting state is
\begin{equation}
\ket{\psi_i(\boldsymbol{\theta})}
=
U_{\mathrm{VQC}}
\left(
\boldsymbol{\theta}
\right)
U_{\mathrm{enc}}
\left(
\mathbf{z}_i
\right)
\ket{\psi_0}.
\label{eq:vqc_state}
\end{equation}
For the binary task, the quantum layer returns Pauli-$Z$ expectation values from two output qubits:
\begin{equation}
m_{i,k}
=
\left\langle
\psi_i(\boldsymbol{\theta})
\middle|
Z_k
\middle|
\psi_i(\boldsymbol{\theta})
\right\rangle,
\qquad
k=1,2,
\label{eq:quantum_measurements}
\end{equation}
where $Z_k$ denotes the Pauli-$Z$ operator acting on output qubit $k$.
The quantum-output vector is therefore
\begin{equation}
\mathbf{m}_i
=
\left[
m_{i,1},m_{i,2}
\right]^{\mathsf{T}}
\in[-1,1]^2.
\label{eq:quantum_output}
\end{equation}
The shallow design limits simulation cost and supports future evaluation under finite-shot and noisy quantum settings. However, the present experiments use noiseless analytic expectation values and exclude finite-shot uncertainty, hardware noise, connectivity, transpilation, and queueing effects.

\subsection{Classical Classification Head}
\label{sec:classification_head}

In the full-hybrid model, the quantum-output vector passes through a trainable classical head to produce two class logits:
\begin{equation}
\mathbf{s}_i
=
f_{\mathrm{head}}
\left(
\mathbf{m}_i;
\boldsymbol{\omega}
\right),
\qquad
\mathbf{s}_i\in\mathbb{R}^{2},
\label{eq:classical_head}
\end{equation}
where $\boldsymbol{\omega}$ denotes the head parameters. The corresponding class probabilities are
\begin{equation}
p_{i,c}
=
\frac{\exp(s_{i,c})}
{\sum_{r=0}^{1}\exp(s_{i,r})},
\qquad
c\in\{0,1\}.
\label{eq:class_probabilities}
\end{equation}
For binary evaluation, the anomaly score is the probability assigned to class $1$:
\begin{equation}
\hat{y}_i^{q}
=
p_{i,1},
\label{eq:hybrid_anomaly_score}
\end{equation}
where class $0$ denotes the normative class and class $1$ denotes the combined suspicious/high-anomaly class. This score feeds \eqref{eq:gamma} and the risk-fusion process and is not interpreted as a calibrated probability of malicious activity.

\subsection{End-to-End Training}
\label{sec:end_to_end_training}

For a mini-batch $\mathcal{B}$, training uses a class-weighted negative
log-likelihood objective:
\begin{equation}
\mathcal{L}_{\mathrm{NLL}}
=
-\frac{1}{|\mathcal{B}|}
\sum_{i\in\mathcal{B}}
w_{y_i}
\log p_{i,y_i},
\label{eq:nll_loss}
\end{equation}
where $w_{y_i}$ denotes the inverse-frequency class weight associated with label $y_i$. The class weights are computed from the training partition and normalized to unit mean. The embedder, VQC, and classification-headparameters are jointly optimized with respect to $\mathcal{L}_{\mathrm{NLL}}$.
The PennyLane quantum node is integrated with the PyTorch computational graph, enabling gradients through all three components. Training uses a learning rate of $2\times10^{-3}$, mini-batches of 32, 20 epochs, and random seeds 42--46. Because circuit outputs use analytic expectation values, the results describe a noiseless simulated hybrid model and exclude shot noise.

\subsection{Hybrid Model Ablations}
\label{sec:hybrid_ablations}

Five configurations evaluate the hybrid components:
\texttt{full\_hybrid} uses the complete embedder, two-layer VQC, and
classification head; \texttt{no\_head} removes the full head;
\texttt{shallow\_embedder} reduces the embedder;
\texttt{no\_embedder} directly encodes the processed features; and
\texttt{pqc\_only} uses a restricted PQC without the complete embedding/head
structure. These ablations assess component contributions under the same
simulated setting without implying quantum computational advantage.

\section{Performance Evaluation}
\label{sec:perform_eval}

We evaluate VQC--ZTI using CESNET-derived aggregated backbone-traffic records and a hybrid PyTorch--PennyLane implementation~\cite{vqc_zti_framework_2026}. The full-hybrid QNN is compared with classical baselines and model ablations under random, entity-group, and temporal holdouts. We also examine the implications for zero-trust enforcement and interpret the prototype's timing within the split-plane architecture.

\subsection{Dataset, Labeling, and Experimental Setup}
\label{sec:dataset_labeling}

The evaluation uses CESNET-derived aggregated flow records~\cite{koumar2025cesnettimeseries24}. The retained payload-independent attributes include packet and byte counts, destination diversity, ratios, average flow duration, and average time-to-live. After validation, missing-value filtering, and feature preparation, the dataset contains
\begin{equation}
N=4{,}875,
\qquad
F=12,
\label{eq:evaluation_dimensions}
\end{equation}
where $N$ and $F$ denote the number of records and features, respectively.
For each record $i$, the anomaly statistic $a_i$ is defined as the Euclidean norm of its robust-scaled feature vector:
\begin{equation}
a_i
=
\left\|
\widetilde{\mathbf{x}}_i
\right\|_2,
\qquad
\widetilde{x}_{i,k}
=
\frac{x_{i,k}-\operatorname{med}_k}
{\operatorname{IQR}_k},
\label{eq:anomaly_statistic}
\end{equation}
where $\operatorname{med}_k$ and $\operatorname{IQR}_k$ denote the median and inter-quartile range of feature $k$, respectively. The configured statistical-anomaly thresholds use the $0.85$ and $0.95$ empirical quantiles of $\{a_i\}$ to distinguish normative, suspicious, and high-anomaly records.
The corresponding empirical boundaries over the processed dataset are
\begin{equation}
\begin{aligned}
b_{\mathrm{susp}}
&=
Q_{0.85}
\left(
\{a_i\}_{i=1}^{N}
\right),\\
b_{\mathrm{high}}
&=
Q_{0.95}
\left(
\{a_i\}_{i=1}^{N}
\right).
\end{aligned}
\label{eq:pseudo_label_quantiles}
\end{equation}
Here, $Q_p(\cdot)$ denotes the empirical $p$-quantile. Records satisfying $a_i<b_{\mathrm{susp}}$ are labeled normative, those satisfying $b_{\mathrm{susp}}\leq a_i<b_{\mathrm{high}}$ are labeled suspicious, and those satisfying $a_i\geq b_{\mathrm{high}}$ are labeled high-anomaly.
For binary benchmarking, the latter two tiers are combined:
\begin{equation}
y_i=
\begin{cases}
0, & \text{normative record},\\
1, & \text{suspicious or high-anomaly record}.
\end{cases}
\label{eq:binary_evaluation_labels}
\end{equation}
These pseudo-labels provide a controlled statistical benchmark rather than verified attack annotations. Accordingly, the reported false-positive rate measures disagreement with the quantile-derived normative class rather than false alarms against verified ground truth.
The labeling quantiles, detector threshold $\gamma^q$, and policy thresholds $\boldsymbol{\tau}$ have distinct roles: the quantiles construct experimental labels, $\gamma^q$ converts anomaly scores into detector decisions, and $\boldsymbol{\tau}$ maps fused risk to enforcement actions.
We use three holdout protocols: \textit{1) random stratified}: 3,900 training and 975 evaluation records, preserving the binary-label distribution; \textit{2) entity-group}: Disjoint entity groups, with 3,751--3,971 training and 904--1,124 evaluation records; and \textit{3) temporal}: 3,887 earlier records for training and 988 later records for evaluation.
The processed features and labels are used consistently across
protocols; model fitting uses only training records, and performance is reported on the corresponding holdout. Each 12-qubit, two-layer QNN is trained for 20 epochs with learning rate $2\times10^{-3}$, batch size 32, and seeds 42--46; results are mean $\pm$ standard deviation across five runs. Analytic expectation values exclude finite-shot and physical-device noise.
The ablations are \texttt{full\_hybrid}, \texttt{no\_head}, \texttt{shallow\_embedder}, \texttt{no\_embedder}, and \texttt{pqc\_only}. Performance is measured using AUC, accuracy, and false-positive rate (FPR):
\begin{equation}
\mathrm{FPR}
=
\frac{\mathrm{FP}}
{\mathrm{FP}+\mathrm{TN}}.
\label{eq:fpr}
\end{equation}
Here, $\mathrm{FP}$ and $\mathrm{TN}$ denote false-positive and true-negative predictions, respectively, relative to the quantile-derived binary labels.

\subsection{Detector Performance Across Splits}
\label{sec:detector_results}

\begin{figure}[ht]
\centering
\begin{minipage}[ht]{0.32\linewidth}
  \centering
  \includegraphics[width=\linewidth]
  {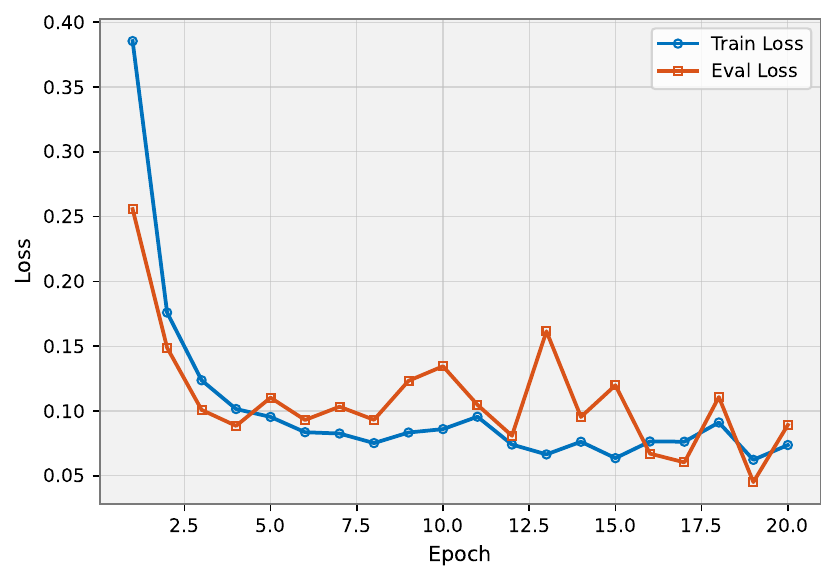}
  \\[-1mm]\footnotesize (a) Training and evaluation loss
\end{minipage}\hfill
\begin{minipage}[ht]{0.32\linewidth}
  \centering
  \includegraphics[width=\linewidth]
  {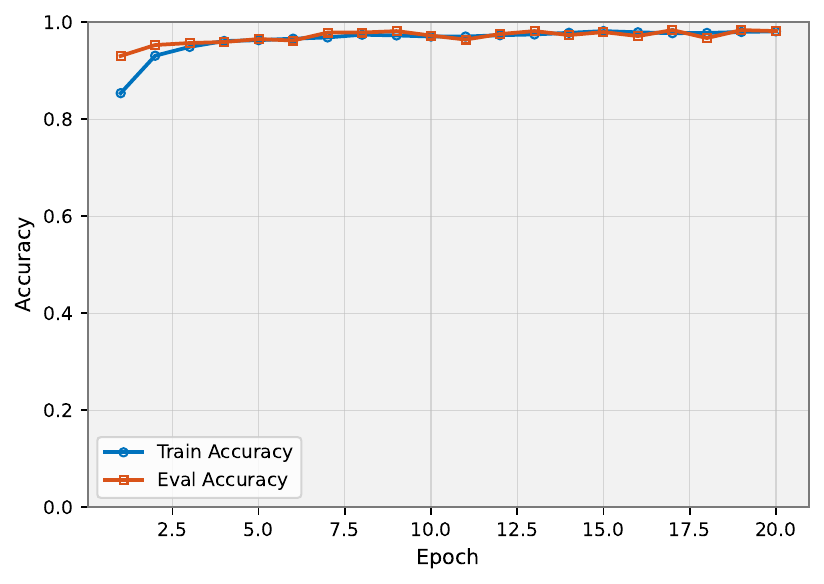}
  \\[-1mm]\footnotesize (b) Training and evaluation accuracy
\end{minipage}\hfill
\begin{minipage}[ht]{0.32\linewidth}
  \centering
  \includegraphics[width=\linewidth]
  {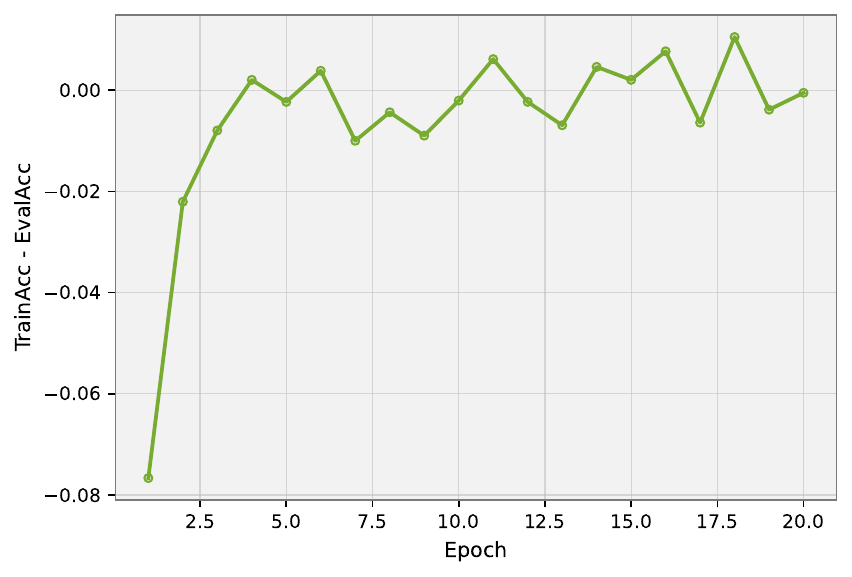}
  \\[-1mm]\footnotesize (c) Training and evaluation gap
\end{minipage}
\caption{Representative full-hybrid QNN training behavior on the random split.}
\label{fig:rand_training_curves}
\end{figure}

Fig.~\ref{fig:rand_training_curves} shows representative full-hybrid QNN training behavior for seed 44 under the random-stratified holdout. The decreasing loss in Fig.~\ref{fig:rand_training_curves}(a) and stabilizing accuracy in Fig.~\ref{fig:rand_training_curves}(b) indicate convergence within 20 epochs. The train--evaluation gap in Fig.~\ref{fig:rand_training_curves}(c) does not show pronounced late-epoch divergence in this representative run. Table~\ref{tab:cross_split_results} provides the primary cross-seed evidence.
The QNN is compared with ExtraTrees (XT), Random Forest (RF), and Logistic Regression (LR). These provide tree-ensemble and linear reference points but are not parameter- or compute-matched to the hybrid neural architecture; therefore, favorable comparisons do not demonstrate quantum advantage. As shown in Table~\ref{tab:cross_split_results}, the QNN achieves the highest mean AUC and accuracy and the lowest mean FPR under all protocols.
\begin{table}[ht]
\centering
\caption{Cross-split detector performance.}
\label{tab:cross_split_results}
\vspace{0.5mm}
\scriptsize
\setlength{\tabcolsep}{2.5pt}
\renewcommand{\arraystretch}{1.05}
\begin{tabular}{@{}llccc@{}}
\toprule
\textbf{Split} & \textbf{Model} & \textbf{AUC} & \textbf{Accuracy} & \textbf{FPR} \\
\midrule
\multirow{4}{*}{Group}
& \textbf{QNN} & $\mathbf{0.9974\pm0.0009}$ & $\mathbf{0.9762\pm0.0028}$ & $\mathbf{0.0241\pm0.0042}$ \\
& XT & $0.9932\pm0.0014$ & $0.9550\pm0.0147$ & $0.0479\pm0.0191$ \\
& RF & $0.9922\pm0.0013$ & $0.9531\pm0.0147$ & $0.0478\pm0.0209$ \\
& LR & $0.9525\pm0.0075$ & $0.8617\pm0.0232$ & $0.1520\pm0.0307$ \\
\midrule
\multirow{4}{*}{Random}
& \textbf{QNN} & $\mathbf{0.9981\pm0.0004}$ & $\mathbf{0.9756\pm0.0042}$ & $\mathbf{0.0270\pm0.0050}$ \\
& XT & $0.9928\pm0.0006$ & $0.9541\pm0.0106$ & $0.0487\pm0.0137$ \\
& RF & $0.9919\pm0.0007$ & $0.9415\pm0.0175$ & $0.0642\pm0.0245$ \\
& LR & $0.9549\pm0.0058$ & $0.8802\pm0.0132$ & $0.1264\pm0.0182$ \\
\midrule
\multirow{4}{*}{Time}
& \textbf{QNN} & $\mathbf{0.9941\pm0.0018}$ & $\mathbf{0.9747\pm0.0044}$ & $\mathbf{0.0248\pm0.0052}$ \\
& XT & $0.9899\pm0.0000$ & $0.9291\pm0.0000$ & $0.0774\pm0.0000$ \\
& RF & $0.9887\pm0.0000$ & $0.9211\pm0.0000$ & $0.0876\pm0.0000$ \\
& LR & $0.9399\pm0.0000$ & $0.8583\pm0.0000$ & $0.1479\pm0.0000$ \\
\bottomrule
\end{tabular}
\end{table}
ExtraTrees is the strongest classical AUC comparator. Relative to XT, the QNN reduces mean FPR from 0.0487 to 0.0270, from 0.0479 to 0.0241, and from 0.0774 to 0.0248 under the random, entity-group, and temporal holdouts, corresponding to reductions of 44.6\%, 49.6\%, and 67.9\%, respectively.
The entity-group and temporal holdouts provide evidence of statistical-anomaly generalization across unseen entity groups and chronological distribution shifts. However, they measure agreement with the constructed statistical-anomaly labels, not detection of previously unseen attacks.

\subsection{Hybrid-Architecture Ablation}
\label{sec:ablation_results}

\begin{figure*}[!t]
\centering
\begin{minipage}[t]{0.49\textwidth}
\centering
\includegraphics[width=\linewidth]
{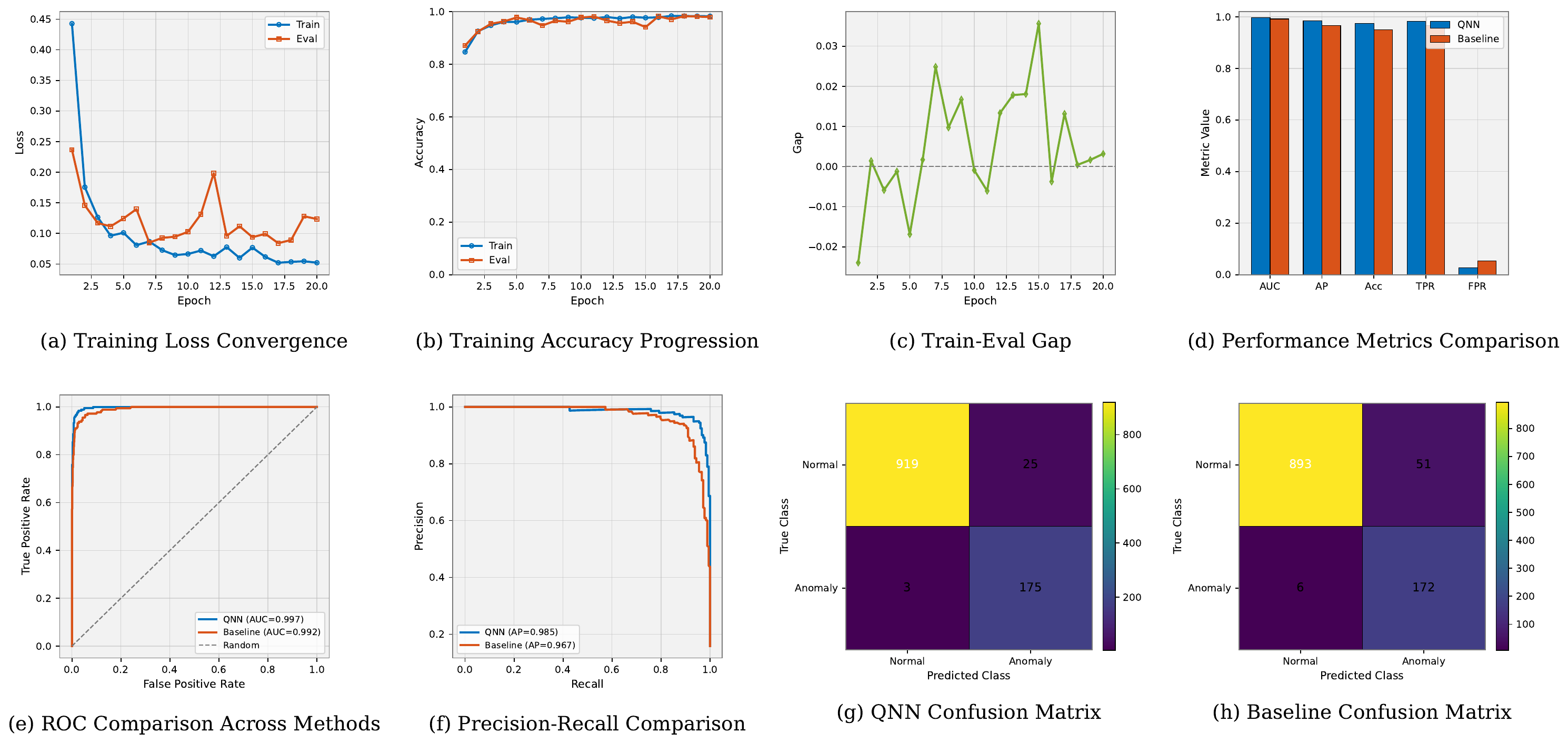}\\[-1mm]
\footnotesize (a) Entity-group holdout
\end{minipage}\hfill
\begin{minipage}[t]{0.49\textwidth}
\centering
\includegraphics[width=\linewidth]
{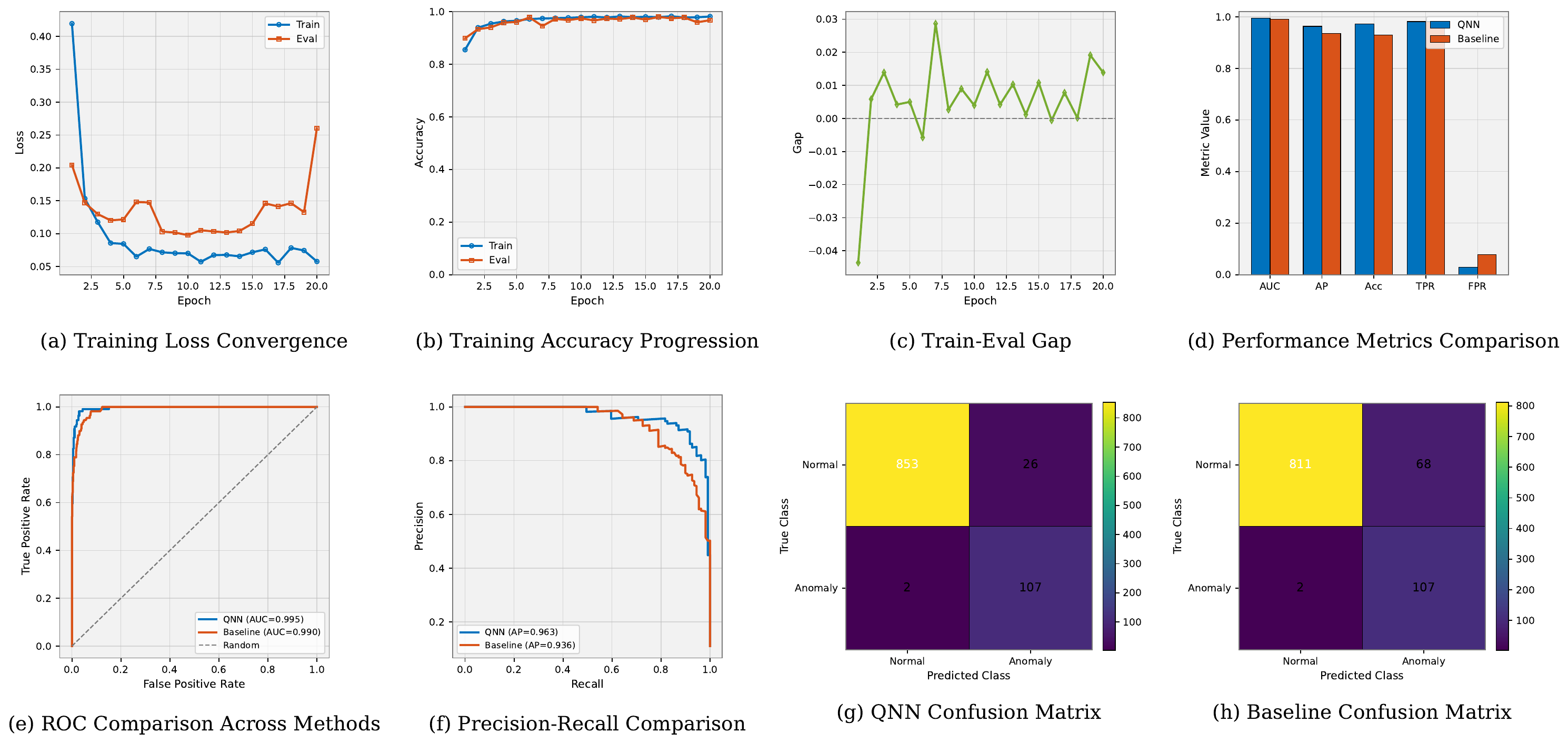}\\[-1mm]
\footnotesize (b) Temporal holdout
\end{minipage}
\caption{Representative full-hybrid QNN and ExtraTrees results under
(a) entity-group and (b) temporal distribution shift.}
\label{fig:shift_summaries}
\vspace{-2mm}
\end{figure*}

Table~\ref{tab:ablation_results} evaluates five variants: Full Hybrid (FH), No Head (NH), Shallow Embedder (SE), No Embedder (NE), and PQC-only (PQC).
Fig.~\ref{fig:shift_summaries} additionally presents representative entity-group and temporal results, while Tables~\ref{tab:cross_split_results} and~\ref{tab:ablation_results} provide the primary cross-seed quantitative evidence.
\begin{table}[ht]
\centering
\caption{Hybrid QNN-model ablation across evaluation splits.}
\label{tab:ablation_results}
\scriptsize
\setlength{\tabcolsep}{2pt}
\renewcommand{\arraystretch}{1.03}
\resizebox{\columnwidth}{!}{
\begin{tabular}{@{}llccc@{}}
\toprule
\textbf{Split} & \textbf{Variant} & \textbf{AUC} & \textbf{Accuracy} & \textbf{FPR} \\
\midrule
\multirow{5}{*}{Group}
& FH  & $\mathbf{0.9974\pm0.0009}$ & $\mathbf{0.9762\pm0.0028}$ & $\mathbf{0.0241\pm0.0042}$ \\
& NH  & $0.9956\pm0.0006$ & $0.9629\pm0.0160$ & $0.0395\pm0.0218$ \\
& SE  & $0.9792\pm0.0065$ & $0.9409\pm0.0088$ & $0.0563\pm0.0111$ \\
& NE  & $0.8693\pm0.0194$ & $0.8071\pm0.0231$ & $0.1922\pm0.0245$ \\
& PQC & $0.8656\pm0.0198$ & $0.8160\pm0.0252$ & $0.1755\pm0.0411$ \\
\midrule
\multirow{5}{*}{Random}
& FH  & $\mathbf{0.9981\pm0.0004}$ & $\mathbf{0.9756\pm0.0042}$ & $\mathbf{0.0270\pm0.0050}$ \\
& NH  & $0.9956\pm0.0013$ & $0.9582\pm0.0068$ & $0.0473\pm0.0081$ \\
& SE  & $0.9770\pm0.0031$ & $0.9175\pm0.0191$ & $0.0869\pm0.0258$ \\
& NE  & $0.8738\pm0.0174$ & $0.8135\pm0.0241$ & $0.1875\pm0.0306$ \\
& PQC & $0.8675\pm0.0209$ & $0.8170\pm0.0410$ & $0.1790\pm0.0545$ \\
\midrule
\multirow{5}{*}{Time}
& FH  & $\mathbf{0.9941\pm0.0018}$ & $\mathbf{0.9747\pm0.0044}$ & $\mathbf{0.0248\pm0.0052}$ \\
& NH  & $0.9927\pm0.0007$ & $0.9640\pm0.0069$ & $0.0371\pm0.0088$ \\
& SE  & $0.9696\pm0.0050$ & $0.9164\pm0.0114$ & $0.0849\pm0.0129$ \\
& NE  & $0.8931\pm0.0081$ & $0.8249\pm0.0230$ & $0.1791\pm0.0307$ \\
& PQC & $0.8888\pm0.0101$ & $0.8553\pm0.0185$ & $0.1413\pm0.0211$ \\
\bottomrule
\end{tabular}
}
\end{table}
The full-hybrid model performs best across all protocols. Removing the classification head causes a modest decline, whereas simplifying or removing the embedder substantially lowers AUC and accuracy and increases FPR. The PQC-only variant also performs materially worse, indicating that the reported performance depends on the complete hybrid pipeline rather than the parameterized quantum circuit alone. The representative results in
Fig.~\ref{fig:shift_summaries} qualitatively reinforce the performance ordering reported in Tables~\ref{tab:cross_split_results} and~\ref{tab:ablation_results}.

\subsection{Implications for Zero-Trust Control}
\label{sec:control_implications}
The detector supplies anomaly evidence to risk fusion rather than issuing access-control commands; combined with identity, device, and segment context, lower disagreement with the normative pseudo-label class may reduce unnecessary restrictions, step-up requests, or denials. However, detector metrics do not establish attack prevention, compromise containment, policy correctness, user disruption, or adversarial service availability, all of which require verified attacks or controlled injections on an implemented PDP/PEP testbed. The statistical labeling tiers, therefore, remain distinct from deployment actions determined by fused contextual risk.
\subsection{Latency and TI-Budget Interpretation}
\label{sec:latency_evaluation}

Table~\ref{tab:prototype_latency} reports prototype-side component timings interpreted within the split-plane architecture rather than as a synchronous VQC decision path. The reported measurements cover the listed software components only; the 0.32-ms VQC scoring value represents an amortized batched measurement and does not include network communication or protected-service processing.
\begin{table}[ht]
\centering
\caption{Split-plane component timings for scoring and enforcement.}
\label{tab:prototype_latency}
\scriptsize
\setlength{\tabcolsep}{2pt}
\renewcommand{\arraystretch}{1.03}
\begin{tabular}{llcc}
\toprule
\textbf{Path} & \textbf{Component} & \textbf{Baseline ZT} & \textbf{VQC--ZTI} \\
\midrule
\multirow{3}{*}{Asynchronous}
& $T_{\mathrm{eval}}$ (Scoring) & 0.85 ms & 0.32 ms (Batched) \\
& $T_{\mathrm{PDP}}$ (Fusion) & 0.25 ms & 0.27 ms \\
& $T_{\mathrm{dist}}$ (Distribution) & 0.40 ms & 0.38 ms \\
\midrule
On-path
& $T_{\mathrm{PEP}}$ (Enforcement) & 0.10 ms & 0.10 ms \\
\bottomrule
\end{tabular}
\end{table}
The 0.32-ms scoring value is an amortized batched measurement rather than single-flow VQC inference time. Anomaly scoring, PDP fusion, and policy distribution belong to the asynchronous update path, whereas cached PEP enforcement lies on the immediate request path. Because communication and protected-service delays were not measured, these component timings do not establish end-to-end TI latency compliance.

\subsection{Limitations and Validity Boundaries}
\label{sec:evaluation_limitations}

The evaluation has four principal limitations: the labels are statistical heuristics rather than verified attack annotations; CESNET flow aggregates do not represent hardware-in-the-loop tactile traffic; noiseless analytic simulation excludes finite-shot and hardware effects; and prototype component timings do not validate end-to-end TI latency. The risk-fusion function and policy thresholds are architectural constructs and were not empirically calibrated against operational access-control outcomes. Verified labels, capacity-matched neural baselines, noisy quantum execution, controlled attacks, and deployment-grade PDP/PEP measurements are therefore required before claiming operational security effectiveness, quantum advantage, or real-time TI compliance.

\section{Conclusions}
\label{sec:conclusions}

VQC--ZTI separates asynchronous hybrid VQC evidence generation from cached, deterministic TI enforcement. On 4,875 CESNET-derived records, the full-hybrid model achieved mean AUCs of 0.9981, 0.9974, and 0.9941 across random, entity-group, and temporal holdouts and reduced mean FPR relative to ExtraTrees by 44.6\%, 49.6\%, and 67.9\%. Ablations show that the complete hybrid architecture, rather than the circuit alone, drives the reported performance.
The results support the feasibility of simulated off-path VQC scoring as a security-evidence source; they do not establish quantum advantage, verified attack detection, hardware execution, or end-to-end TI latency compliance. Future work will use verified attack labels, matched classical neural baselines, finite-shot noisy execution, physical devices, and a hardware-in-the-loop PDP/PEP testbed with tail-latency
and policy-freshness measurements.

\bibliographystyle{IEEEtran}
\bibliography{IEEEabrv,references/refs}

@article{ahmed2025qnnztf,
  title={Quantum-driven zero trust architecture with dynamic anomaly detection in 7G technology: A neural network approach},
  author={Ahmed, Shakil and Shihab, I. F. and Khokhar, Ashfaq},
  journal={Measurement: Digitalization},
  volume={2},
  pages={100005},
  year={2025},
  publisher={Elsevier}
}

@article{promwongsa2021tactile, 
    author = {Promwongsa, Nattakorn and Ebrahimzadeh, Amin and Naboulsi, Diala and Kianpisheh, Somayeh and Belqasmi, Fatna and Glitho, Roch and Crespi, Noel and Alfandi, Omar},
    title = {A Comprehensive Survey of the Tactile Internet: State-of-the-Art and Research Directions}, journal = {IEEE Communications Surveys \& Tutorials},
    volume = {23},
    number = {1},
    pages = {472--523},
    year = {2021}
}

@techreport{rose2020zta,
  author       = {Rose, Scott and Borchert, Oliver and Mitchell, Stu and Connelly, Sean},
  title        = {Zero Trust Architecture},
  institution  = {National Institute of Standards and Technology},
  type         = {Special Publication},
  number       = {800-207},
  year         = {2020}
}

@techreport{chandramouli2023ztaModel,
  author       = {Chandramouli, Ramaswamy and Butcher, Zachary},
  title        = {A Zero Trust Architecture Model for Access Control in Cloud-Native Applications in Multi-Location Environments},
  institution  = {National Institute of Standards and Technology},
  type         = {Special Publication},
  number       = {800-207A},
  year         = {2023}
}

@article{popovski2018urllc,
  author       = {Popovski, Petar and Nielsen, Jimmy J. and Stefanovic, Cedomir and
                  {de Carvalho}, Elisabeth and Str{\"o}m, Erik and
                  Trillingsgaard, Kasper F. and Bana, Alexandru-Sabin and
                  Kim, Dong Min and Kotaba, Radoslaw and Park, Jihong and
                  S{\o}rensen, Ren{\'e} B.},
  title        = {Wireless Access for {URLLC}: Principles and Building Blocks},
  journal      = {IEEE Network},
  volume       = {32},
  number       = {2},
  pages        = {16--23},
  year         = {2018}
}

@article{fettweis2014tactile,
  author       = {Fettweis, Gerhard P.},
  title        = {The Tactile Internet: Applications and Challenges},
  journal      = {IEEE Vehicular Technology Magazine},
  volume       = {9},
  number       = {1},
  pages        = {64--70},
  year         = {2014}
}

@article{durisi2016shortpackets,
  author       = {Durisi, Giuseppe and Koch, Tobias and Popovski, Petar},
  title        = {Toward Massive, Ultra-Reliable, and Low-Latency Wireless Communication with Short Packets},
  journal      = {Proceedings of the IEEE},
  volume       = {104},
  number       = {9},
  pages        = {1711--1726},
  year         = {2016}
}

@article{maier2016tactile,
  author       = {Maier, Martin and Chowdhury, Mahfuzulhoq and Rimal, Bhaskar Prasad and Van, Dung Pham},
  title        = {The Tactile Internet: Vision, Recent Progress, and Open Challenges},
  journal      = {IEEE Communications Magazine},
  volume       = {54},
  number       = {5},
  pages        = {138--145},
  year         = {2016}
}

@article{koumar2025cesnettimeseries24,
  author       = {Koumar, Josef and Hynek, Karel and {\v{C}}ejka, Tom{\'a}{\v{s}} and {\v{S}}i{\v{s}}ka, Pavel},
  title        = {CESNET-TimeSeries24: Time Series Dataset for Network Traffic Anomaly Detection and Forecasting},
  journal      = {Scientific Data},
  volume       = {12},
  number       = {1},
  pages        = {338},
  year         = {2025}
}

@inproceedings{sommer2010outside,
  author       = {Sommer, Robin and Paxson, Vern},
  title        = {Outside the Closed World: On Using Machine Learning for Network Intrusion Detection},
  booktitle    = {Proc. IEEE Symp. Secur. Privacy},
  pages        = {305--316},
  year         = {2010}
}

@article{ring2019datasets,
  author       = {Ring, Markus and Wunderlich, Sarah and Scheuring, Dominic and Landes, Dieter and Hotho, Andreas},
  title        = {A Survey of Network-Based Intrusion Detection Data Sets},
  journal      = {Computers \& Security},
  volume       = {86},
  pages        = {147--167},
  year         = {2019}
}

@article{preskill2018nisq,
  author       = {Preskill, John},
  title        = {Quantum Computing in the {NISQ} Era and Beyond},
  journal      = {Quantum},
  volume       = {2},
  pages        = {79},
  year         = {2018}
}

@article{biamonte2017qml,
  author       = {Biamonte, Jacob and Wittek, Peter and Pancotti, Nicola and Rebentrost, Patrick and Wiebe, Nathan and Lloyd, Seth},
  title        = {Quantum Machine Learning},
  journal      = {Nature},
  volume       = {549},
  pages        = {195--202},
  year         = {2017}
}

@article{mitarai2018qcl,
  author       = {Mitarai, Kosuke and Negoro, Makoto and Kitagawa, Masahiro and Fujii, Keisuke},
  title        = {Quantum Circuit Learning},
  journal      = {Physical Review A},
  volume       = {98},
  number       = {3},
  pages        = {032309},
  year         = {2018}
}

@article{schuld2020circuit,
  author       = {Schuld, Maria and Bocharov, Alex and Svore, Krysta M. and Wiebe, Nathan},
  title        = {Circuit-Centric Quantum Classifiers},
  journal      = {Physical Review A},
  volume       = {101},
  number       = {3},
  pages        = {032308},
  year         = {2020}
}

@misc{vqc_zti_framework_2026,
  author       = {Shakil Ahmed and Mubassir Serneabat Sudipto and Ashfaq Khokhar},
  title        = {VQC-ZTI Framework: Variational Quantum-Classical Zero-Trust Anomaly Detection and CESNET-Based Security Evaluation},
  year         = {2026},
  howpublished = {\url{https://github.com/msudipto/VQC-ZTI_Framework}},
  note         = {Code repository}
}

\end{document}